# Lithography-Free Patterning of $SrTiO_3$ -based Two-Dimensional Electron Gases using Direct Atomic Layer Processing

Anshu Gupta[1], Karolis Parfeniukas[2], Amit Chanda[1], Thor Hvid-Olsen[1], Mira Baraket[2],

Maksym Plakhotnyuk[2], Kasper S. Pedersen[3], Felix Trier[1, *]

[1]Department of Energy Conversion and Storage, Technical University of Denmark, Kgs. Lyngby, Denmark

[2]ATLANT 3D, Taastrup, Denmark

[3]Department of Chemistry, Technical University of Denmark, Kgs. Lyngby, Denmark

Email: fetri@dtu.dk

**Abstract**

We present a scalable and lithography-free strategy for the realization of a two-dimensional electron gas (2DEG) in $TiO_2$-patterned $SrTiO_3$ (100) (STO) via Al deposition using magnetron sputtering. A 15 nm thick $TiO_2$ layer, deposited by direct atomic layer processing (DALP), is employed to spatially define the conducting regions, enabling direct transport measurements without post-growth microfabrication. Upon Al deposition, an insulating $AlO_x$ overlayer is formed, and the region lacking $TiO_2$ pattern leads to the creation of oxygen vacancies in $SrTiO_3$. These oxygen vacancies act as electron donors, populating the Ti 3*d* conduction bands and giving rise to a confined 2DEG at the interface. Magneto-transport measurements reveal a sheet carrier density on the order of $\approx$ 5-7 $\times$ $10^{13}$ $cm^{-2}$, comparable to values typically achieved in pulsed laser deposition-grown $SrTiO_3$-based heterostructures, along with effective electrostatic tunability. This work demonstrates a simple, cost-effective, and industry-compatible route for engineering oxide 2DEGs, providing a versatile platform for scalable device fabrication and interfacial transport studies.

Since 2004, the discovery of an electron gas confined in two-dimensions at the heterointerface between $SrTiO_3$ (STO) and other oxides has attracted substantial recognition due to its emergent physical properties and potential for device applications.[1,2] Two-dimensional electron gases (2DEGs) at STO-based interfaces exhibit fascinating phenomena including coexistence of superconductivity and ferromagnetism[3], Quantum Hall effect[4,5], strong Rashba spin-orbit coupling[6,7], tunability with external stimuli such as light illumination and electrostatic gating[8–11], neuromorphic properties[12], and even signatures of magnetic ordering[13], making them an ideal platform to explore correlated and spin-dependent transport phenomena in low dimensions. In addition to their fundamental relevance, these interfaces have been integrated into functional devices, including oxide transistors[14], rewritable photodetectors, and spin-charge conversion units tailored for advanced spin-based technologies[15,16].

Conventionally, 2DEGs in oxides are realized using epitaxial growth techniques such as pulsed laser deposition (PLD)[17,18] or molecular beam epitaxy (MBE)[19], which demand meticulous control conditions like surface termination, controlled oxygen partial pressure, and high substrate temperature. These growth techniques provide high-quality interfaces having high carrier density but involve complex processing steps and limited scalability. Additionally, for magneto-transport measurements, lithographic patterning involving photoresists is required for defining device geometries which increase fabrication complexity and cost. Developing simpler and cost-effective fabrication approaches with minimal processing complexity, while still maintaining the essential electronic properties, remains a significant challenge in the field. An alternative route to generate 2DEG in STO is based on redox-driven surface reduction[20,21]. Oxidizable metals deposited on STO can extract oxygen from the lattice, forming an insulating metal oxide overlayer and creating oxygen vacancies near the surface[22,23]. These oxygen vacancies act as electron donors, populating Ti 3*d* conduction bands and resulting in a confined electron gas[24,25].

In this work, we demonstrate the lithography-free formation of a two-dimensional electron gas (2DEG) through the deposition of metallic Al by DC magnetron sputtering onto $TiO_2$-patterned $SrTiO_3$ (STO) substrates. The $TiO_2$ patterns are fabricated using Direct Atomic Layer Processing (DALP), a spatially selective variant of atomic layer deposition (ALD) that enables localized material growth without the need for photoresists, masks, or post-deposition patterning steps. Unlike conventional oxide-electronic device fabrication, where transport structures are typically defined through photolithography, resist processing, etching, and lift-off steps, DALP provides direct-write patterning of functional oxide layers in a single additive process. This eliminates chemical contamination associated with photoresists, reduces process complexity, and enables rapid design iteration through software-defined patterning. While the present demonstration employs feature sizes in the hundreds-of-micrometers regime, the approach is inherently compatible with further scaling and high-throughput manufacturing.

In DALP, precursor gases are delivered through a confined microreactor positioned in close proximity to the substrate surface, allowing the self-limiting surface reactions characteristic of ALD to occur only within predefined regions. By controlling the relative motion between the substrate and the reactor head, arbitrary patterns can be deposited directly with sub-nanometer thickness precision while maintaining the conformality, uniformity, and film quality associated with conventional ALD processes.[26,27] The deposited $TiO_2$ layer serves as a spatially selective redox-blocking mask, enabling direct definition of the active conducting regions without conventional lithographic processing. Furthermore, the self-limiting nature of ALD chemistry ensures excellent thickness reproducibility, uniformity, and process scalability, while the software-defined patterning approach enables straightforward modification of device geometries without the need for redesigning photomasks. These attributes make DALP an attractive platform for the fabrication of oxide-electronic devices and for exploring lithography-free routes toward complex oxide heterostructures. Electrical transport experiments reveal that the resulting 2DEG shows metallic behavior upon cooling from 300 K to 2 K with a reduction of the sheet resistance from 11.5 kΩ/sq. to 0.7 kΩ/sq. Magneto-transport measurements at 2 K reveal a sheet carrier density $n_s \approx 6.2 \times 10^{13}$ cm$^{-2}$ well within the range of electrostatic gate-tunable carrier densities and an electron mobility $\mu \approx 867.5$ cm$^2$ V$^{-1}$ s$^{-1}$. Notably, the obtained carrier density is comparable to that reported for STO-based oxide heterostructures fabricated using PLD. Our approach offers a simpler and more scalable approach compared to PLD and MBE based growth techniques, eliminating the need for complex deposition conditions and conventional lithographic processes. These results highlight the potential of sputter-deposited metal/oxide heterostructures as a cost effective and scalable platform for realizing 2DEGs in oxide electronics and next generation device applications.

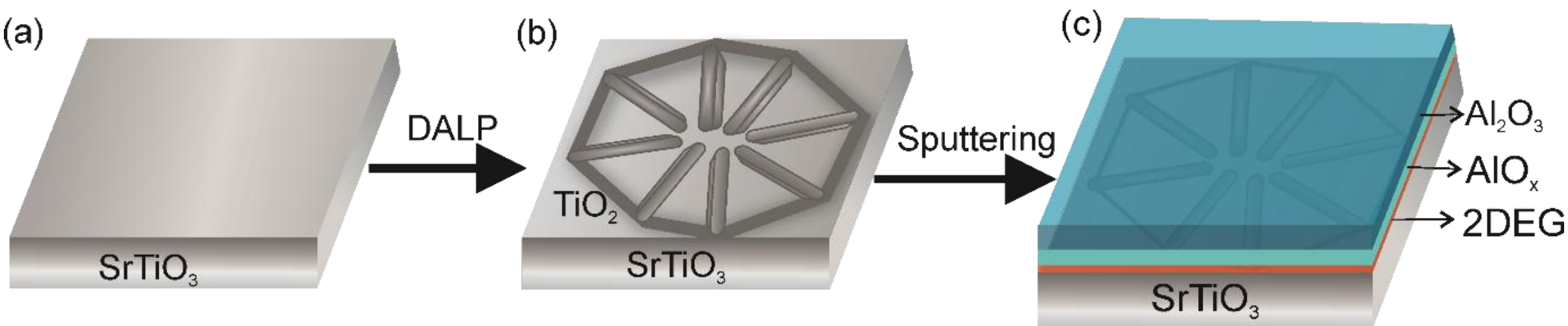

***Figure 1:*** Schematic illustration of the fabrication process. (a) As-received $SrTiO_3$ (100) substrate, annealed at 1000 °C for 1 hour in 1 bar of $O_2$. (b) Patterned $TiO_2$ layer deposited by direct atomic layer processing (DALP) by ATLANT 3D. (c) Deposition of Al by DC magnetron sputtering, leading to surface reduction of STO and formation of a 2DEG. The Al-induced 2DEG is subsequently capped with an $Al_2O_3$ layer for stabilization and protection.

Figure 1 illustrates the fabrication sequence of the heterostructure starting with commercial as-received STO (100) substrates, serving as the platform for the 2DEG formation. The STO substrates were annealed in 1 bar $O_2$ atmosphere at 1000 °C for 1 hour with a heating/cooling ramp rate of 100 °C/hour. After annealing, a 15 nm thick $TiO_2$ layer is selectively deposited and patterned using direct atomic layer processing (DALP) based on the atomic layer deposition (ALD) principle by ATLANT 3D (see Fig. 1(b)). A thickness of 15 nm was selected to ensure complete suppression of oxygen diffusion and redox interaction with the underlying STO during subsequent Al deposition, while maintaining mechanical stability and sharp lateral pattern definition. The $TiO_2$ thin film was deposited using a well-known TTIP and water reaction[28] at 150 °C substrate temperature. The thickness-controlled nature of ALD with accuracy down to sub-nm precision ensures a uniform and well defined $TiO_2$ structure with sharp pattern boundaries. The DALP tool used in this study enables lateral patterning with a precision on the order of a few hundred micrometers, sufficient for the proof-of-concept transport devices demonstrated here. Although conventional UV and electron-beam lithography can achieve substantially smaller feature sizes, DALP provides a lithography-free and resist-free patterning approach with excellent thickness control and process reproducibility. Furthermore, ongoing improvements in DALP hardware are expected to enable lateral resolutions below 100 μm. This patterned $TiO_2$ plays a crucial role in spatially confining the conducting regions, as it selectively blocks or permits the redox driven 2DEG formation at the STO surface[22]. The patterned geometry directly defines the active device area, eliminating the need for complex post-growth lithography and enabling precise control over device architecture. Initially, a thin Al layer was deposited directly onto the STO substrate using DC magnetron sputtering. The substrate temperature was 150 °C throughout the growth process (see Experimental details). During deposition, a redox reaction occurs between Al and oxygen atoms from the STO surface, leading to the formation of an insulating $AlO_x$ layer. This reaction creates oxygen vacancies at the STO surface, which act as electron donors. The released electrons occupy the Ti 3*d* conduction bands, thereby realizing a 2DEG at the interface[29]. Subsequently, without breaking the vacuum, an $Al_2O_3$ protective capping layer was deposited *in situ* using RF magnetron sputtering to prevent environmental degradation of the STO-based 2DEG by dioxygen[30].

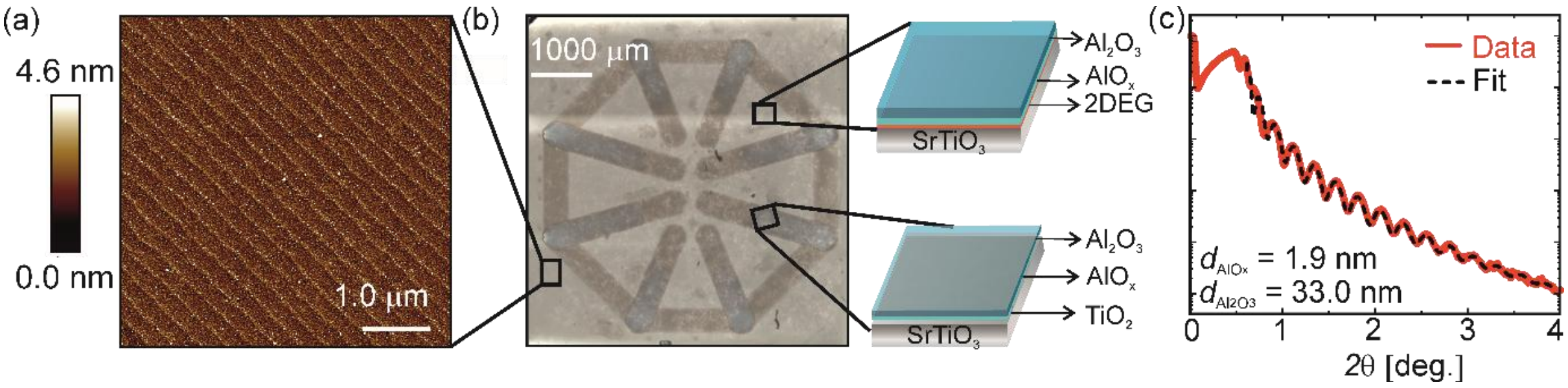

***Figure 2.*** (a) Atomic force microscopy images of an $Al_2O_3/AlO_x/TiO_2$-STO heterostructure with root mean square (RMS) roughness of 0.5 nm. (b) Optical microscopy image of the heterostructure $Al_2O_3/AlO_x/TiO_2$-STO. (c) Kiessig fringes in the X-ray reflectivity (XRR) data for $Al_2O_3/AlO_x/TiO_2$-STO.

Figure 2(b) displays the optical microscopy image of the patterned device fabricated on a STO substrate. The darker, radially arranged structure corresponds to the patterned $TiO_2$ mask deposited by DALP on the STO surface ($TiO_2$-STO). The symmetric geometry enables multi-terminal transport measurements with a defined current path and voltage probe configuration. The unpatterned regions exhibited clear formation of a two-dimensional electron gas (2DEG), whereas no 2DEG formation was observed in the patterned areas of the STO substrate. This is illustrated by two schematics in the magnified view. The X-ray diffraction data for the bare $TiO_2$-STO substrate (before Al deposition) and $Al_2O_3/AlO_x/$ $TiO_2$-STO are shown in Supplementary Fig. S1. Figure 2(c) shows the Kiessig fringes in the X-ray reflectivity (XRR) spectrum associated with a thick $Al_2O_3$ capping layer on the $AlO_x$ layer. By fitting the XRR data, the thickness of the $AlO_x$ film was found to be $1.9 \pm 0.9$ nm and the $Al_2O_3$ layer thickness was determined to $33.0 \pm 0.9$ nm. Before Al deposition, the morphological properties of the $TiO_2$-STO substrate were studied by atomic force microscopy (AFM). The surface shows an atomically clean step-terrace structure with a 0.4 nm height (Fig. S2) and a low surface roughness of 0.3 nm. The AFM image obtained after the deposition of $AlO_x$ and $Al_2O_3$ shows a similar topography as the STO surface, with a slightly increased surface roughness of 0.5 nm (Figure 2(a)).

Figure 3(a) displays the employed geometry for the magneto-transport measurement of the 2DEG formed at the $AlO_x$/STO interface. The grey bar indicates the length ($L$) and width ($W$) of the conduction channel, required for the calculation of the sheet resistance ($R_s$) defined as $R_s = R_{xx}\ W/L$, where $R_{xx}$ is the longitudinal resistance given as $V_{xx}/I$. The temperature ($T$) dependence of $R_s$ reveals the presence of metallic behavior over the measured temperature range from 300 K to 5 K as shown in Fig. 3 (b), with a residual resistivity ratio of $R_s$(300 K)/$R_s$(5 K) = 16.5. However, an upturn in the resistance below $T$ = 20 K can be observed, which is generally attributed to Kondo-like screening, resulting from exchange interactions between itinerant electrons of the 2DEG and with localized magnetic defects. The experimental $R_s$(T) data was fitted in this interface in the range 5 K - 40 K using the following general expression[31–33]:

$$R_s\,(T) = R_o + R_{inel}\,T^n + \; R_{Kondo}\,\{1 - \left(\frac{\ln(T/_{T_{k^*}})}{\sqrt{\{\ln(T/_{T_{k^*}})\}^2 + \pi^2 S(S+1)}}\right)\} \tag{1}$$

Here, $R_o$ represents the temperature-independent residual resistance, the second term is accounting for the combined electron-electron inelastic and electron-phonon scattering. In equation (1), the third term represents the Kondo-like scattering contribution, where $T_k^*$ is the effective Kondo temperature and $S$ is the effective spin of the magnetic scattering centers, which was fixed at $S$ = 0.225[34]. The fitting parameters are listed in Table S1. The effective Kondo temperature was $T_k^* = 13.9 \pm 1.5$ K as obtained from the fit. The evolution of the longitudinal resistance with magnetic field, defined as magnetoresistance (MR), was investigated by applying an out-of-plane magnetic field between +9 T and –9 T. The MR is defined as MR = ($R_s$ ($B$) – $R_s$ ($B$=0)) /$R_s$ ($B$=0). For the $AlO_x$/STO interface, a positive MR is observed over the entire temperature range, and the absolute value of MR changed from 0.1% at 300 K to 10.5 % at 2 K. A weak low-field cusp in the magnetoresistance is observed at 2 K, which may be indicative of weak anti-localization (WAL) arising from spin-orbit coupling in the STO-based 2DEG. However, the feature is relatively weak and becomes absent for $T \geq 5$ K and higher temperatures. We note that the low-temperature magneto transport may also be influenced by other scattering mechanisms, such as Kondo scattering. The maximum value of MR of approximately 10.7% ($B$ = ± 9 T) was found to occur at $T$ = 10 K. With increasing temperature, the magnitude of the MR gradually decreases, and it becomes nearly negligible at room temperature as shown in Figure 3(c). This non-monotonic temperature dependence is consistent with the presence of a competing scattering mechanism at low temperatures. We attribute this behavior to Kondo-like scattering arising from localized magnetic moments or defect states, which compete with weak antilocalization two band contributions to transport.

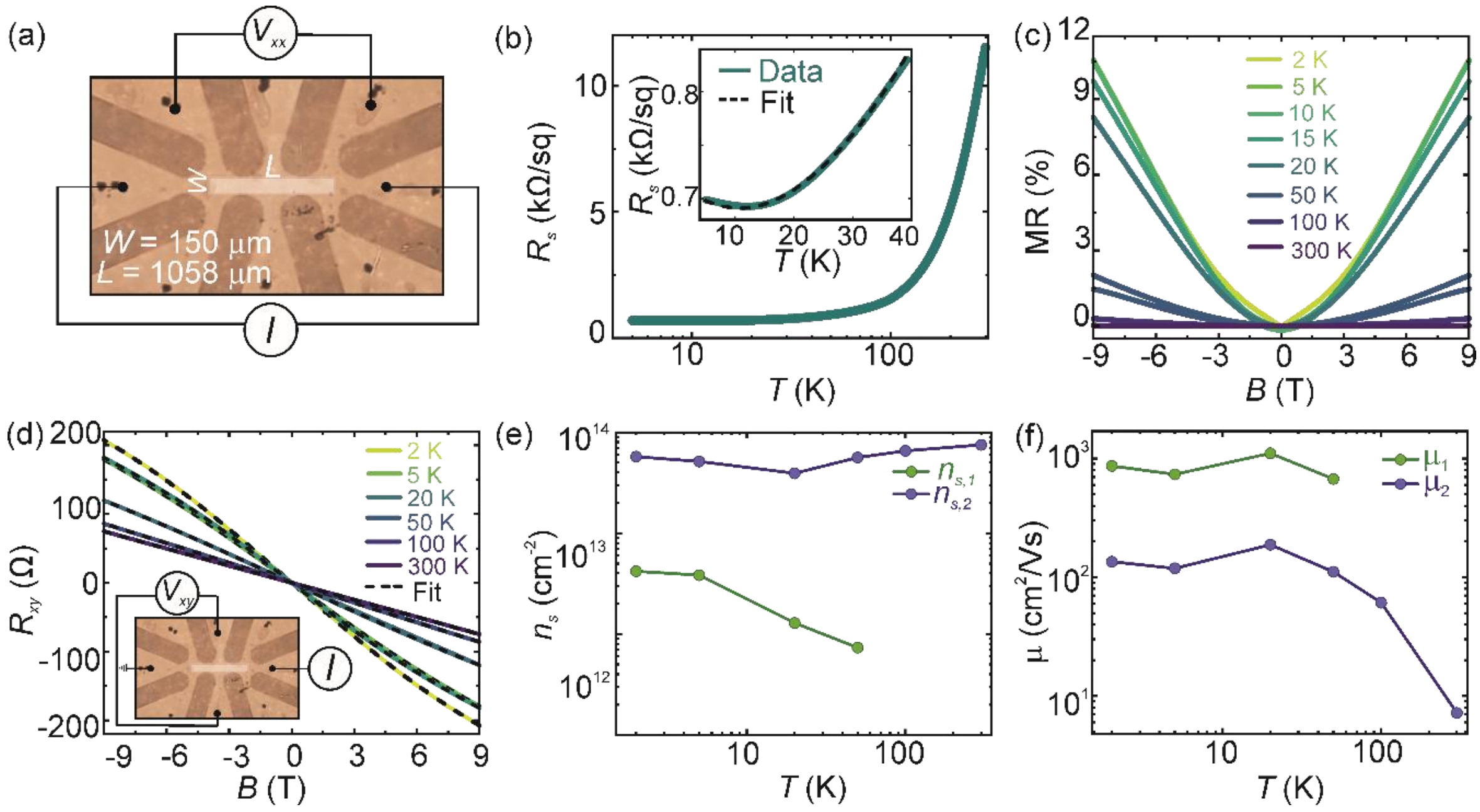

***Figure 3***. (a) Optical microscopy image of the heterostructure with the longitudinal contacts for magneto-transport measurement. The blue bar shows the length (*L*) and width (*W*) of the conduction channel. (b) Temperature dependence of the sheet resistance, $R_s(T)$ between 5 K and 300 K. The inset shows the fitting of the low temperature upturn in $R_s(T)$ using Eq. (1). (c) Magnetic field dependence of the magnetoresistance (MR = $[R_s(B) - R_s(B=0)\}/ R_s(B=0)]$) obtained while sweeping an out-of-plane magnetic field between $B = \pm 9$ T at 2 K, 5 K, 10 K, 15 K, 20 K, 50 K, 100 K, and 300 K. (d) Hall resistance ($R_{xy}$) of the 2DEGs in the heterostructure $Al_2O_3/AlO_x/TiO_2$-STO, measured from $B = -9$ T to +9 T at different temperatures. The inset shows the geometry for the Hall measurement. Temperature dependent variation of (e) the charge carrier densities $n_s$, and (f) charge carrier mobilities $\mu$. The error bars in (e) & (f) are incorporated in the marker size.

Further, to determine the charge carrier density and mobility of the 2DEG formed at the interface, Hall measurements were performed over a temperature range (2-300 K) for the magnetic field range $B$ = –9 T to +9 T. Figure 3(d) shows the Hall resistance $R_{xy} = V_{xy}/I$ as a function of $B$ at various temperatures, where $V_{xy}$ is the Hall voltage and $I$ the sourcing current (fixed to 100 μA). The inset illustrates the transverse contact configuration used for the Hall measurements. Below 100 K, $R_{xy}$ exhibits a pronounced non-linear dependence with the applied magnetic field ($B$), whereas for $T$ > 100 K, a linear $R_{xy}(B)$ behavior is observed. As commonly observed in STO-based 2DEGs, the non-linearity of the field-dependence of the low-temperature Hall resistance is consistent with the presence of two types of charge carriers contributing to the transport properties[23,35,36]. In contrast, the linear behavior at higher temperatures suggests a dominant single-carrier transport regime. The charge carrier densities ($n_s$) and mobilities ($\mu$) of the individual conduction channels were extracted by fitting the experimental Hall resistance data to a classical two-band model[36,37].

$$R_{xy}(B) = -\left(\frac{1}{e}\right)\left(\frac{\left\{\frac{n_{s,1}\mu_1^2}{1+\mu_1^2B^2}+\frac{n_{s,2}\mu_2^2}{1+\mu_2^2B^2}\right\}.B}{\left\{\frac{n_{s,1}\mu_1}{1+\mu_1^2B^2}+\frac{n_{s,2}\mu_1}{1+\mu_2^2B^2}\right\}^2+\left\{\frac{n_{s,1}\mu_1^2}{1+\mu_1^2B^2}+\frac{n_{s,2}\mu_2^2}{1+\mu_2^2B^2}\right\}^2.B^2}\right) \quad (2)$$

where $n_{s,1}$ and $n_{s,2}$ denote, the densities of the two species of charge carriers, and $\mu_1$, $\mu_2$ are the corresponding Hall mobilities. Here, the equation (2) is constrained by the expression

$$R_s^{-1}\,(B=0) = e\,(n_{s,1}\mu_1 + n_{s,2}\mu_2) \quad (3)$$

The temperature evolution of the charge carrier density and the mobility is shown in Figure 3(e) and (f). As temperature decreases from 300 K to 5 K, the transport evolves from a single carrier regime to a two-carrier conduction mechanism. At 300 K, only one type of charge carrier is present, with the total sheet carrier density of $n_{s,2} \approx 7.5 \pm 0.0 \times 10^{13}$ cm$^{-2}$. Upon cooling below 100 K, a second carrier channel emerges. At 2 K, the carrier density $n_{s,1}$ increases to $4.2 \pm 0.1\times 10^{12}$ cm$^{-2}$, while the second carrier density $n_{s,2}$ decreases to $5.7 \pm 0.1 \times 10^{13}$ cm$^{-2}$.

Correspondingly, the carrier mobility shows a significant enhancement upon cooling, where $\mu_1$ reaches ≈ 867.5 ± 2.3 $cm^2\ V^{-1}\ s^{-1}$ (2 K), whereas the mobility of the second carrier type ($\mu_2$) exhibits an increase from ≈ 6.7 ± 0.0 (at 300 K) to 135.3 ± 1.9 $cm^2\ V^{-1}\ s^{-1}$ (2 K). Overall, the total sheet carrier density $n_s$ decreases from 7.5 ± 0.0 × $10^{13}$ to 6.2 ± 0.1 x $10^{13}$ $cm^{-2}$. This charge carrier density is similar to the other oxide STO heterostructures forming 2DEGs via sputtering and PLD[35,38–43]. The mobility on the order of ~ 867.5 ± 2.3 $cm^2\ V^{-1}\ s^{-1}$ at 2 K is consistent with the sheet carrier density (~5.7 ± 0.1 × $10^{13}$ $cm^{-2}$). This behavior can be attributed to enhanced scattering with ionized donor atoms, in line with established scattering models[44]. However, by reducing unintentional impurities, lattice defects and interface disorder the mobility can be further improved by optimized growth conditions. Importantly, this obtained mobility represents no significant trade-off over established growth methods like PLD and MBE despite the significantly simplified processing, lower thermal budget, and lithography-free pattern definition.

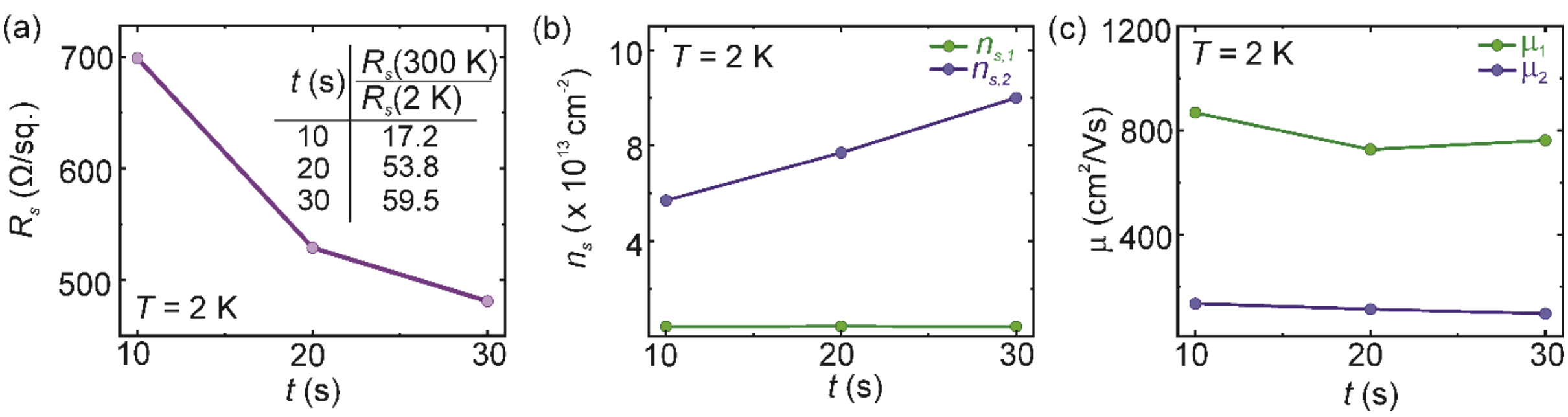

Figure 4: (a) Sheet resistance ($R_s$) measured at 2 K as a function of Al deposition time ($t$) during sputtering, with the substrate temperature ($T_G$) maintained at 150 °C. The inset shows the ratio of $R_s$ measured at 300 K and 2 K. (b) Charge carrier concentration as a function of depending on Al deposition time at 2 K. (c) Carrier mobility as a function of Al deposition time at 2 K.

Figure 4(a) represents the dependence of $R_s$ measured at 2 K as a function of Al deposition time during sputtering of samples prepared with $TiO_2$-STO substrates. $R_s$ decreases systematically with increasing deposition time, corresponding to a greater $AlO_x$ thickness. Specifically, $R_s$ decreases from approximately ≈698.8 Ω/sq. for a 10 s deposition time to ≈480.0 Ω/sq. for a 30 s deposition. The inset table shows the residual resistivity ratio ($R_s$ (300 K)/ $R_s$ (2 K)) for the samples grown at 150 °C showing increase in the ratio with Al deposition time. This reduction in sheet resistance indicates that the charge carrier density can be effectively tuned by controlling the deposited metal thickness[20]. The optical microscopy images and measurement configurations for 20 s and 30 s deposited samples are shown in Supplementary Fig. S4.
Figure 4 (b) indicates the variation of charge carrier density ($n_s$) with Al deposition time at 2 K. The dominant charge carriers ($n_{s,2}$) increases from 5.7 × $10^{13}$ $cm^{-2}$ for 10 s deposition time to 1.0 × $10^{14}$ $cm^{-2}$ for 30 s deposition time, while $n_{s,1}$ remains nearly constant. The corresponding mobility shows a gradual decrease with increasing carrier density, as expected from enhanced scattering (Fig. 4 (c)). These results collectively demonstrate a reproducible thickness-dependent evolution of transport properties. The transport data for these samples at 300 K are provided in Supplementary Table 2. The dependence of $R_s$ measured at 300 K as a function of Al deposition time during sputtering of samples prepared without $TiO_2$ mask is shown in supplementary figure S5.

In summary, we demonstrate a simple and proof-of-principle route to realize a 2DEG in $TiO_2$-patterned STO by depositing Al using magnetron sputtering. This approach avoids complex oxide epitaxy and eliminates the need for conventional lithography, as the $TiO_2$ patterning directly defines active regions suitable for device architecture. The resulting 2DEG exhibits a total sheet carrier density on the order of 6.2 × $10^{13}$ $cm^{-2}$, comparable to values reported for PLD-grown STO based heterostructures. The use of Direct Atomic Layer Processing (DALP) for $TiO_2$ pattern definition is central to this approach, as it enables high-quality, thickness-controlled, and spatially selective redox blocking without the need for resist-based lithography. The intrinsic material precision and lateral programmability of DALP provide a scalable pathway toward more complex oxide device architectures with improved spatial resolution. These findings show that sputtering-based metal/oxide architecture provides a practical and industry-compatible platform for engineering oxide 2DEGs, offering strong potential for scalable device fabrication and further studies of interfacial electronic phenomena, including aperiodic quantum oscillations[45], charge to spin current conversion[46], non-reciprocal charge transport[47,48] and the quantum metric of electrons[49].

Prior to the deposition, the $TiO_2$-patterned STO substrates were first ultrasonicated in acetone and then in isopropanol, each for 8 mins. Initially, a thin Al layer was deposited directly onto the STO substrate using DC magnetron sputtering at a working pressure of $3.3 \times 10^{-3}$ mbar and a DC power of 40 W. The substrate temperature was 150 °C throughout the growth process. The target-to-substrate distance was fixed at 25 mm, and the deposition was carried out for 10 s in an argon atmosphere without intentional oxygen flow. The $Al_2O_3$ deposition was performed at a working pressure of $3.5 \times 10^{-3}$ mbar with an RF power of 100 W, a target-to-substrate distance of 45 mm, and a deposition time of 10 min.

The structural properties of $Al_2O_3/AlO_x/TiO_2$-STO were characterized with XRD and XRR measurements using a Rigaku Smart Lab diffractometer, where XRR data were analyzed using a multilayer model fitting approach with the built-in genetic algorithm in the Rigaku SmartLab Studio software, where the thickness, density, and interfacial roughness of each layer were used as adjustable fitting parameters. The surface morphological properties were characterized with Bruker AFM. The four-terminal electrical measurements on this heterostructure were performed using a Quantum Design Physical Properties Measurement System (by Quantum Design) in the magnetic field range of $B = \pm 9$ T, with temperature varying from $T = 2$ K to 300 K, using the DC resistivity option. The longitudinal resistance ($R_{xx}$) and Hall resistance ($R_{xy}$) were processed by performing symmetrization and antisymmetrization respectively, to eliminate mixing effects arising from contact misalignment and experimental asymmetries. Specifically, $R_{xx}^{sym}$ was obtained from symmetric part by doing symmetrization

$$R_{xx}^{sym}(B) = \frac{R_{xx}(B) + R_{xx}(-B)}{2}$$

while $R_{xy}^{asym}$ was extracted from the antisymmetric part,

$$R_{xy}^{asym}(B) = \frac{R_{xy}(B) - R_{xy}(-B)}{2}$$

The raw data, along with the symmetrized magnetoresistance (MR) and antisymmetrized Hall resistance data, have been included in Supplementary Figure S6. The electrical contacts at this interface 2DEG in this $AlO_x/TiO_2$-STO were made by ultrasonic Al wire wedge bonder without needing any extra contact pads.

**Supplementary Material:** The supplementary material provides additional experimental details, characterization data, and analysis supporting the findings presented in this work.

**Acknowledgments**

A. G., A. C., T. H. O. and F. T. acknowledge support by research grant 37338 (SANSIT) from Villum Fonden. A.C. and F. T acknowledge the support from Villum Fonden (ETHOS, Grant No. 69171).

**Conflicts of interest**

The authors declare no conflict of interest.

**Data Availability statement**

The data that supports findings of this study are available from the corresponding author upon reasonable request.

**References**

[1] A. Ohtomo, and H.Y. Hwang, "A high-mobility electron gas at the LaAlO3 /SrTiO3 heterointerface," Nature **427**(1), 229–233 (2025).

[2] J. Levy, "Physics of SrTiO3 -based heterostructures and nanostructures," Rep. Prog. Phys. **81**, 036503 (2018).

[3] L. Li, C. Richter, J. Mannhart, and R.C. Ashoori, "Coexistence of magnetic order and two-dimensional superconductivity at LaAlO3 / SrTiO3 interfaces," Nat. Phys. **7**(9), 762–766 (2011).

[4] Y. Matsubara, K.S. Takahashi, M.S. Bahramy, Y. Kozuka, D. Maryenko, J. Falson, A. Tsukazaki, Y. Tokura, and M. Kawasaki, "Observation of the quantum Hall effect in δ-doped SrTiO3," Nat. Commun. **7**(5), (2016).

[5] F. Trier, G.E.D.K. Prawiroatmodjo, Z. Zhong, D.V. Christensen, M. Von Soosten, A. Bhowmik, J.M.G. Lastra, Y. Chen, T.S. Jespersen, and N. Pryds, "Quantization of Hall Resistance at the Metallic Interface between an Oxide Insulator and SrTiO3," Phys. Rev. Lett. **117**(9), (2016).

[6] W. Lin, L. Li, F. Do, C. Li, H. Rotella, X. Yu, B. Zhang, Y. Li, W.S. Lew, S. Wang, W. Prellier, S.J. Pennycook, J. Chen, Z. Zhong, A. Manchon, and T. Wu, "Interface-based tuning of Rashba spin-orbit interaction in asymmetric oxide heterostructures with 3d electrons," Nat. Commun. **10**, 1–7 (2019).

[7] A.D. Caviglia, M. Gabay, S. Gariglio, N. Reyren, C. Cancellieri, and J. Triscone, "Tunable Rashba Spin-Orbit Interaction at Oxide Interfaces," Phys. Rev. Lett. **126803**(3), 1–4 (2010).

[8] K. Ueno, S. Nakamura, H. Shimotani, A. Ohtomo, N. Kimura, T. Nojima, H. Aoki, Y. Iwasa, and M. Kawasaki, "Electric-field-induced superconductivity in an insulator," Nat. Mater. **7**, 855–858 (2008).

[9] J.T. Ye, S. Inoue, K. Kobayashi, Y. Kasahara, H.T. Yuan, H. Shimotani, and Y. Iwasa, "Liquid-gated interface superconductivity on an atomically flat film," Nat. Mater. **9**(11),

125–128 (2010).

[10] S. Zhang, Y. Lin, C. Nan, R. Zhao, and J. He, “Magnetic and Electrical Properties of (Mn, La)-Codoped SrTiO3 Thin Films,” **3266**, (2008).

[11] D.V. Christensen, F. Trier, W. Niu, Y. Gan, Y. Zhang, T.S. Jespersen, Y. Chen, and N. Pryds, “Stimulating Oxide Heterostructures : A Review on Controlling SrTiO3 -Based Heterointerfaces with External Stimuli,” Adv. Mater. Interfaces **1900772**, 1–40 (2019).

[12] X. Yan, X. Han, Z. Fang, Z. Zhao, Z. Zhang, Z. Guo, X. Jia, Y. Zhang, Z. Guan, and T. Shi, “Reconfigurable memristor based on SrTiO3 thin-film for neuromorphic computing,” Front. Phys. **18**, 63301 (2023).

[13] A. Brinkman, M. Huijben, M. Van Zalk, J. Huijben, U. Zeitler, J.C. Maan, W.G.V. der A.N.D.E.R. Wiel, G. Rijnders, D.H.A. Blank, and H. Hilgenkamp, “Magnetic effects at the interface between non-magnetic oxides,” Nat. Mater. **6**, 493–496 (2007).

[14] S. Datta, and B. Das, “Electronic analog of the electro-optic modulator,” Appl. Phys. Lett. **56**, 665–667 (1990).

[15] P. Noël, F. Trier, L.M.V. Arche, J. Bréhin, D.C. Vaz, V. Garcia, S. Fusil, A. Barthélémy, L. Vila, M. Bibes, and J. Attané, “Non-volatile electric control of spin – charge conversion in a SrTiO3 Rashba system,” Nature **580**(August 2019), (2020).

[16] F. Gallego, F. Trier, S. Mallik, J. Bréhin, S. Varotto, L. Moreno Vicente-Arche, T. Gosavy, C.-C. Lin, J.-R. Coudevylle, L. Iglesias, and others, “All-Electrical Detection of the Spin-Charge Conversion in Nanodevices Based on SrTiO3 2-D Electron Gases,” Adv. Funct. Mater. **34**(3), 2307474 (2024).

[17] M.L. Scullin, J. Ravichandran, C. Yu, and M. Huijben, “Pulsed laser deposition-induced reduction of SrTiO3 crystals,” Acta Mater. **58**(2), 457–463 (2010).

[18] T. Ohnishi, H. Koinuma, and M. Lippmaa, “Pulsed laser deposition of oxide thin films,” Appl. Surf. Sci. **252**, 2466–2471 (2006).

[19] D.G. Schlom, “Perspective: Oxide molecular-beam epitaxy rocks!,” APL Mater. **3**(7), 062403 (2016).

[20] L.M. Vicente-arche, S. Mallik, M. Cosset-cheneau, P. Noël, D.C. Vaz, F. Trier, T.A. Gosavi, C. Lin, D.E. Nikonov, I.A. Young, A. Sander, A. Barthélémy, J. Attané, L. Vila, and M. Bibes, “Metal / SrTiO3 two-dimensional electron gases for spin-to-charge conversion,” Phys. Rev. Mater. **064005**, 2–11 (2021).

[21] T.C. Rödel, F. Fortuna, S. Sengupta, E. Frantzeskakis, P. Le Fèvre, F. Bertran, B. Mercey, S. Matzen, G. Agnus, T. Maroutian, P. Lecoeur, and A.F. Santander-Syro, “Universal Fabrication of 2D Electron Systems in Functional Oxides,” Adv. Mater. **28**(10), 1976–1980 (2016).

[22] T.C. Rodel, F. Fortuna, S. Sengupta, E. Frantzeskakis, P. Le F, B. Mercey, S. Matzen, G. Agnus, T. Maroutian, and P. Lecoeur, “Universal Fabrication of Two-Dimensional Electron Systems in Functional Oxides,” Adv. Mater. **1980**, 1976–1980 (2016).

[23] Y. Chen, N. Pryds, E. Kleibeuker, G. Koster, J. Sun, and E. Stamate, “Metallic and Insulating Interfaces of Amorphous SrTiO3 -Based Oxide Heterostructures,” Nano Lett. **11**(9), 3774–3778 (2011).

[24] X.H. Huang, Z. Huang, S.W. Zeng, X.P. Qiu, L.S. Huang, A. Annadi, J.S. Chen, J.M.D. Coey, and T. Venkatesan, “Origin of the Two-Dimensional Electron Gas at LaAlO3- SrTiO3 Interfaces : The Role of Oxygen Vacancies and Electronic Reconstruction,” Phys. Rev. X **3**, 021010(1–9) (2013).

[25] J. Delahaye, and T. Grenet, “Metallicity of the SrTiO3 surface induced by room temperature evaporation of alumina,” J. Phys. D Appl. Physics, **45**, 315301 (2012).

[26] I. Kundrata, M.K.S. Barr, S. Tymek, D. Döhler, B. Hudec, P. Brüner, G. Vanko, M. Precner, T. Yokosawa, E. Spiecker, M. Plakhotnyuk, K. Fröhlich, and J. Bachmann, “Additive Manufacturing in Atomic Layer Processing Mode,” Small Methods **6**, 2101546

(2022).

[27] S. Santucci, A. Kinikar, Z. Wang, N.S. Ruhela, J. Navne, M. Akbari, A. Mishchenko, M. Baraket, and M. Plakhotnyuk, "Growth dynamics in patterned TiO2 deposited by direct atomic layer processing (DALP) in ambient conditions," J. Vac. Sci. Technol. A **43**, 062411 (2025).

[28] B.A. Rahtu, and M. Ritala, "Reaction Mechanism Studies on Titanium Isopropoxide - Water Atomic Layer Deposition Process**," Chem. Vap. Depos. **8**, 21–28 (2002).

[29] H. Xu, Y. Gan, Y. Zhao, M. Li, X. Hu, X. Chen, R. Wang, Y. Li, J. Sun, F. Hu, Y. Chen, and B. Shen, "Two-Dimensional Electron Gases at the Amorphous and Crystalline SrTiO3/KTaO3 Heterointerfaces," Phys. Status Solidi Appl. Mater. Sci. **220**(13), (2023).

[30] F. Trier, D. V Christensen, Y.Z. Chen, A. Smith, M.I. Andersen, and N. Pryds, "Degradation of the interfacial conductivity in LaAlO3 / SrTiO3 heterostructures during storage at controlled environments," Solid State Ionics **230**, 12–15 (2013).

[31] S. Qi, H. Zhang, J. Zhang, Y. Gan, X. Chen, B. Shen, Y. Chen, Y. Chen, and J. Sun, "Large Optical Tunability of 5d 2D Electron Gas at the Spinel / Perovskite γ -Al2O3 / KTaO3 Heterointerface," Adv. Mater. Interfaces **2200103**, 3–9 (2022).

[32] M. Lee, J.R. Williams, S. Zhang, C.D. Frisbie, and D. Goldhaber-Gordon, "Electrolyte gate-controlled Kondo effect in SrTiO3," Phys. Rev. Lett. **107**(25), 256601 (2011).

[33] H. Zhang, H. Zhang, X. Yan, X. Zhang, Q. Zhang, J. Zhang, F. Han, L. Gu, B. Liu, Y. Chen, B. Shen, and J. Sun, "Highly Mobile Two-Dimensional Electron Gases with a Strong Gating E ff ect at the Amorphous LaAlO3 / KTaO3 Interface," ACS Appl. Mater. Interfaces **9**, 36456–36461 (2017).

[34] M. Lee, J.R. Williams, S. Zhang, and C.D. Frisbie, "Electrolyte Gate-Controlled Kondo Effect in SrTiO3," **256601**(12), 1–5 (2011).

[35] A. Joshua, S. Pecker, J. Ruhman, E. Altman, and S. Ilani, "A universal critical density

underlying the physics of electrons at the LaAlO3/SrTiO3 interface," Nature **3**, 1–7 (2012).
[36] J.S. Kim, S.S.A. Seo, M.F. Chisholm, R.K. Kremer, H. Habermeier, B. Keimer, and H.N. Lee, "Nonlinear Hall effect and multichannel conduction in LaTiO3- SrTiO3 superlattices," Phys. Rev. B **82**, 2–5 (2010).
[37] F. Gunkel, C. Bell, H. Inoue, B. Kim, A.G. Swartz, and T.A. Merz, "Defect Control of Conventional and Anomalous Electron Transport at Complex Oxide Interfaces," Phys. Rev. X **031035**, 1–15 (2016).
[38] C.R. N. Reyren, S. Thiel, A. D. Caviglia, L. Fitting Kourkoutis, G. Hammerl, D.A.M. C. W. Schneider, T. Kopp, A.-S. Rüetschi, D. Jaccard, M. Gabay, and J.M. J.-M. Triscone, "Superconducting Interfaces Between Insulating Oxides," Science **317**(8), 1196–1200 (2007).
[39] G.E.D.K. Prawiroatmodjo, F. Trier, D. V Christensen, Y. Chen, N. Pryds, and T.S. Jespersen, "Evidence of weak superconductivity at the room-temperature grown LaAlO3 / SrTiO3 interface," Phys. Rev. B **184504**, 2–6 (2016).
[40] A.D. Caviglia, S. Gariglio, N. Reyren, D. Jaccard, T. Schneider, M. Gabay, S. Thiel, G. Hammerl, J. Mannhart, and J. Triscone, "Electric field control of the LaAlO3 / SrTiO3 interface ground state," Nature **456**(12), 2–5 (2008).
[41] C. Bell, S. Harashima, Y. Kozuka, M. Kim, B.G. Kim, Y. Hikita, and H.Y. Hwang, "Dominant Mobility Modulation by the Electric Field Effect at the LaAlO3 -SrTiO3 Interface," Phys. Rev. Lett. **226802**(11), 25–28 (2009).
[42] D.A. Dikin, M. Mehta, C.W. Bark, C.M. Folkman, C.B. Eom, and V. Chandrasekhar, "Coexistence of Superconductivity and Ferromagnetism in Two Dimensions," Phys. Rev. Lett. **056802**(7), 1–4 (2011).
[43] S. Caprara, and J. Lesueur, "Multiple quantum criticality in a two-dimensional superconductor," Nat. Mater. **12**(6), 542–548 (2013).

[44] F. Trier, D. V Christensen, and N. Pryds, "Electron mobility in oxide heterostructures," J. Phys. D. Appl. Phys. **51**, 293002 (2018).

[45] K. Rubi, J. Gosteau, R. Serra, K. Han, S. Zeng, Z. Huang, B. Warot-Fonrose, R. Arras, E. Snoeck, Ariando, M. Goiran, and W. Escoffier, "Aperiodic quantum oscillations in the two-dimensional electron gas at the LaAlO3/SrTiO3 interface," Npj Quantum Mater. **5**(1), (2020).

[46] J. Bréhin, L.M.V. Arche, S. Varotto, S. Mallik, J. Attané, L. Vila, A. Barthélémy, N. Bergeal, and M. Bibes, "Gate-voltage switching of nonreciprocal transport in oxide-based Rashba interfaces," Phys. Rev. Appl. **10**(1), 1 (2023).

[47] D. Choe, M.J. Jin, S.I. Kim, H.J. Choi, J. Jo, I. Oh, J. Park, H. Jin, H.C. Koo, B.C. Min, S. Hong, H.W. Lee, S.H. Baek, and J.W. Yoo, "Gate-tunable giant nonreciprocal charge transport in noncentrosymmetric oxide interfaces," Nat. Commun. **10**(1), 1–8 (2019).

[48] D.C. Vaz, F. Trier, A. Dyrdał, A. Johansson, K. Garcia, A. Barthélémy, I. Mertig, and J. Barna, "Determining the Rashba parameter from the bilinear magnetoresistance response in a two-dimensional electron gas," Phys. Rev. Mater. **071001**, 1–7 (2020).

[49] G. Sala, M.T. Mercaldo, K. Domi, S. Gariglio, M. Cuoco, and C. Ortix, "The quantum metric of electrons with spin-momentum locking," Science **389**, 822–825 (2025).